\documentstyle{article}
\author{Gorsky O.I. \and Kuchugurny Yu.P.}
\title{ Long-time unbreaking correlations in the large number
        of degrees of freedom Hamiltonian system}
\date{}
\begin{document}
\maketitle

\begin{abstract}
A behaviour of molecular cluster with Lennard-Jones potential of interactions
as Hamiltonian system is studied by computer simulation (molecular dynamics
method). It is shown that complex periodic oscillations of the cluster
as a whole are possible. This is in accordance with KAM theorem.
\end{abstract}

\section{Introduction}

We study  the behaviour of the dynamical system
with Lennard - Jones interaction by computer simulation.
The main physical motivation is concerned with long - time
dynamical evolution in nonlinear Hamiltonian systems perturbed
by random fluctuations caused by rounding-off errors.

The goal of works [1-4] was the investigation of a dynamical
foundations of classical statistical mechanics.
It were considered :\\
{  - } a chain of $ N $ nonlinearly coupled particles whose Hamiltonian
is given by [3] ( Fermi - Pasta - Ulam $\beta$ - model )

\begin{equation}
H = \sum_{i=1}^N \left( \frac{1}{2} \dot \phi_{i}^{2}
+\frac{1}{2}(\phi_i - \phi_{i+1})
+ \frac{1}{4} \beta (\phi_i -\phi_{i+1})^{4} \right)
\end{equation}

where $\phi_i$ are the displacement with respect to stable positions.\\
{  - }a chain of $N$ nonlinearly coupled particles with Hamiltonian [3]

\begin{equation}
H = \sum_{i=1}^N \left( \frac{1}{2} \dot \phi_{i}^{2}+ \frac{1}{2}(\phi_i
- \phi_{i+1})
+ \frac{\alpha}{3} ( \phi_i -\phi_{i+1})^3 \right)
\end{equation}

{  - }the system of $N$ particles arranged on a square lattice with nonlinear
interaction through a Lennard - Jones potential [1]

\begin{equation}
V(r) = 4\epsilon \left(
A_1\left(\frac{\sigma}{r}\right)^{12}
-A_2 \left(\frac{\sigma}{r}\right)^6
\right)
\end{equation}

$\epsilon$ is a depth of the potential well;\\
$\sigma =r_0 2^{-1/6}$, where $V(r_0) = 0 $, in [1-2]
were taken $A_1=1,A_2=1$ .\\
Boundary conditions have been chosen periodic in normal mode coordinates [3]
$$
\phi_i = \phi_{i+N} ,
$$
or square lattice was surrounded by a border of fixed particles [1-2].

In all cited works it have been proposed that particles oscillate
about their stable positions. The main results there obtained
may be summarized as follows:\\
{  - }at low energy there exist ordered motions, normal modes appear to be
uncoupled [2]; \\
{  - }correspondengly to a certain value of the average - time kinetic energy
per particle relaxation toward the equilibrium is strongly slowed down [4];\\
{  - }a time of relaxation $\tau_r(\epsilon)$ ( $\epsilon$ is a full energy per
particle in normal modes terms ) is compatible with Nekhoroshev - like
low [4]  with model dependent parameter $\delta$

\begin{equation}
\tau_r = \tau_0 \exp ( (\frac{\epsilon_0}{\epsilon})^\delta )
\end{equation}

At high energy the transition between ordered - disordered ( weakly -
strongly chaotic dynamics ) occurs but there are no exist thresholds
between two dynamical regimes. In all cited papers the choice of an
indicator of system's state was found to be insufficient. The use
of relaxation time $\tau_r(\epsilon)$ is initial conditions dependent
[4], spectral entropy $\eta(t)$ is different from zero when
$ t \to \infty $ [3-4], Laypounov characreristic exponent $\lambda(\epsilon)$
is initial conditions independent but gives model - dependent
transition  and poor convergence when number of particles $N$ is large.

The main physical conclusion is the following:
" there is a nonvanishing measure of initial conditions
in phase space that can originate long living metastable states even
in the thermodynamical limit " [4].

It has been studied Hamiltonian dynamics subjected to rounding - off errors
fluctuations. The tests concerned the Hamiltonian dynamics were
the following: negligibly small value of
$\Delta \epsilon / \epsilon \simeq 10^{-5} - 10^{-6} $
and time reversible dynamics for $ t \to -t $ [4].

Based on cited results it may be concluded that:\\
{  - }with the proviso that particles oscillate close to positions where
their potential energy has a minimum, provided that  number of
the positions equal to  number of particles , when full enrgy
$\epsilon > \epsilon_c$ ( $\epsilon_c$ is model and initial conditions
dependent) a number of oscillations with continuous random spectrum
may increase on account of decrease in oscillations
with discrete spectrum;\\
{  - }apparently in general way a characteristic time
$\tau_{r}^{(1)}(\epsilon)$ for breaking of discrete spectrum is not existent;\\
{  - }a finite characteristic time $\tau_{r}^{(2)}(\epsilon)$
within which one would to determine  a random nature of oscillations in system
is not existent;\\
{  - }there is a fundamental problem of long - time following of  dynamics
in computer simulations because of rounding - off errors. This are nearly
tangent to problem of modelling of dynamical evolution.

The starting positions of the work based on assumption
that weak changes of oscillations modes in dynamical system with large
number of degrees of freedom may give rise to complex behavior of whole
system when to abandon any restriction in motions and in particular one could
to observe the dynamics without any changes in oscillations modes during
long - time computer simulations with unbreaking correlations
of the motions in a whole system despite of rounding - off errors.

\section{Model description}

Computer simulation runs have been performed for $N$ particles Lennard -Jones
system without any boundary or the like conditions. The equations of motion
have been computed with undimensional time - step
$\Delta t = 0.05 - 0.13$, where real time - step
\begin{equation}
\Delta t_{real} =
\Delta t \left(\frac{m\sigma^2}{48\epsilon}\right)^{1/2}
\end{equation}
The undimensional average - time kinetic energy per particle
is defined as follow

\begin{equation}
\langle\epsilon_{kin}\rangle =
\frac{1}{N} \sum_{i=1}^{N}\langle\left( v_{xi}^2+v_{yi}^2 \right)\rangle
\end{equation}

Computer simulations were carried out without any cut off range of $r$.

Initial conditions are of particular significance in present work.
The starting points of particles are setting up as follows

\begin{equation}
x_i = \alpha A j \sin \left(\frac{2\pi i}{N_i}\right),
y_i =        A j \cos \left(\frac{2\pi i}{N_i}\right)
\end{equation}

\noindent
$
A =6.752/4.53,
                N_i = 22 \mbox{ when } j = 4, i =1...22,
\mbox{ and }    N_i = 32 \mbox{ when } j = 5,\\ i =1...32,
\mbox{ and }    N_i = 44 \mbox{ when } j = 6, i =1...44,
\alpha = 1.013, A_1 = 4.870, A_2 = 1.
$

At the first stage of formation of the cluster next scaling way was used.
After time $ t= 200 k \Delta t , k=1...5,$ over the particles velocities
were set $ v_{xi} = 0, v_{yi} = 0 $ for $i =1...N$.
After some time $t_1$  the scaling of velocities was abandoned
and the cluster became as isolated but subjected to rounding - off errors.
So the dynamical system is  Hamiltonian or near Hamiltonian
as fluctuations of full energy $\Delta \epsilon / \epsilon$ were
of the order of $ 10^{-4}- 10^{-5}$ and average - time
$\langle \Delta \epsilon \rangle$ was equal to zero.

\section{Results description}

The graphical displays of the cluster during program runs is depicted
in Fig.1. The cluster oscillates as whole with period $ t^* \simeq 2000 $
and is unbreaking by rounding - off errors ( in this case the time of the
unbreaking correlations of motion of cluster as a whole was $t_r=5\quad 10^4 $,
time step $\Delta t =0.13$). The dependence of time of an unbreaking correlations
vs its full energy per particle is reported in Fig.2
( $t_1 =1100 $ in this case ).
In Fig.3 is shown the value of average kinetic energy per particle.
It is displayed time - average kinetic energy $\langle\epsilon_{kin}\rangle$ though all particles of
the cluster has a complex periodic oscillations about some state ( Fig.4).

The cluster doesn't lose this complex periodic oscillation when time - step
of integration was 2 times increased  and doesn't lose its symmetry when
program runs ( $t_r = 10^5 $ ,time - step $\Delta t=0.13$) in the case
when full energy per particle is about 0.01 (in undimensional units).

The number of particles with similar symmetrical behavior is no less
then two - fold.

\section{Discussion}

In accordance with KAM theorem such
cluster with complex periodic oscillations is possible.

What is to bring out the existence in computer simulation such cluster?\\
{  - } the cluster has sensitivity to initial conditions and to parameters
 $ A_1, A_2 $ of potential of interaction. All of this are characteristic
for chaotic system. \\
{  - } the cluster brings out the unbreaking collective correlations.  The
average - time kinetic energy equal to 1-3 K ( parameters $\epsilon,
\sigma$ were taken for argon ).
This is the interval of an energy where convergence of normal modes
in questionable in accordance with [1-2].\\
{  - } collective correlations of the motion increase the stability
of the cluster as a whole despite of rounding - off errors.

Is the cluster artificial or has its physical reality ? This is out
of the purpose of the present work. The dynamics of the cluster
is near Hamiltonian with $ \langle \epsilon \rangle \simeq const $.

We would like to underline that symmetry in initial position
of the particles brings  complex periodic oscillations in this case.
For the cluster with parameters
$
A =6.752/4.53,
N_i = 20 \mbox{ when } j = 4, i = 1...20,
N_i = 30 \mbox{ when } j = 5, i = 1...30,
N_i = 40 \mbox{ when } j = 6, i = 1...40,
\alpha = 1.000, A_1 = 4.870, A_2 = 1.
$
one could to observe spontaneous turning of symmetry axis induced
by rounding - off errors. Spontaneous breaking or turning of symmetry
of any kind in Hamiltonian dynamics is impossible.

Some words about description parameters of the cluster where one could
observe contraction - extention of the cluster as a whole.
When time - space scale is large then one needs only 4 parameters
to describe contraction - extention of the cluster as a whole.
For molecular space and time scale one needs much more number
of degrees of desription's parameters ( but less then $4N$).

There is main question from the work. We would like to formulate
this question as follow:
What number of degrees of freedom, full energy and potential of interaction
did provide time of an unbreaking correlations in a cluster subjected
to rounding - off errors for an infinity or this time is always finite ?

\medskip
\noindent
Address:
Transmag Research Institute,
Piesarjevsky Str. 5,
Dniepropetrovsk, 320005
Ukraine\\
e-mail:root@tmg.dp.ua
\medskip

\end{document}